\documentclass[
prd 
,showpacs, amssymb, superscriptaddress, aps
]{revtex4}
\usepackage{graphicx}
\input{epsf}

\usepackage{amsmath,amssymb}
\usepackage{bm}
\usepackage{times}

\newcommand{\dalm}{\kern1pt\vbox{\hrule height 0.9pt\hbox{\vrule width 0.9pt
\hskip 2.5pt\vbox{\vskip 5.5pt}\hskip 3pt\vrule width 0.3pt}\hrule height 0.3pt}
\kern1pt}

\newcommand{\gsim}{\, \raisebox{-0.8ex}{$\stackrel{\textstyle >}{\sim}$ }}
\newcommand{\lsim}{\, \, \raisebox{-0.8ex}{$\stackrel{\textstyle <}{\sim}$ }}

\newcommand{\be}{\begin{eqnarray}}
\newcommand{\ee}{\end{eqnarray}}
\newcommand{\beq}{\begin{eqnarray}}
\newcommand{\eeq}{\end{eqnarray}}

\newcommand{\ie}{i.e.,~}
\newcommand{\eg}{e.g.,~}

\usepackage{color}

\definecolor{maroon}{cmyk}{0,0.87,0.68,0.32}

\usepackage{graphicx}

\begin{document}



\title{Compactness of neutron stars and Tolman VII solutions in scalar-tensor gravity}

\author{Hajime Sotani}
\email{sotani@yukawa.kyoto-u.ac.jp}
\affiliation{Division of Theoretical Astronomy, National Astronomical Observatory of Japan, 2-21-1 Osawa, Mitaka, Tokyo 181-8588, Japan}

\author{Kostas D. Kokkotas}
\affiliation{Theoretical Astrophysics, IAAT, Eberhard-Karls University of T\"{u}bingen, 72076 T\"{u}bingen, Germany}

\date{\today}

\begin{abstract}
We systematically examine the compactness of neutron stars as Tolman VII solutions in scalar-tensor theory of gravity. As a result, when the coupling constant is confined to values provided by astronomical observations we show that the maximum compactness of neutron stars in general relativity is higher than that in scalar-tensor gravity. In addition, we show that although ultra-compact stars, with radius smaller than the Regge-Wheeler potential peak, can exist in general relativity (\eg Tolman VII solution), their scalarized counterparts cannot {be constructed} even in the limiting case of  uniform density stars.
\end{abstract}

%

\maketitle

\section{Introduction}
\label{sec:I}

General relativity is  the most reliable theory for describing gravity. In fact, up to now, astronomical observations and terrestrial experiments are  in remarkable good agreement with the predictions in general relativity. Even so, most of tests have been performed in the weak-field regime, \eg in our solar system. This means that the observational verifications of general relativity in the strong-field regime are still poor, and the true theory of gravity in these region may deviate from general relativity. In such a case, some strong-field phenomena  may deviate from the predictions in general relativity and one may be able to probe the gravitational theory by measuring and monitoring these potential deviations.  In practice, several possibilities for observationally probing the gravitational theory have already been proposed (e.g., \cite{Berti2015,DP2003,SK2004,Sotani2014}). In any case, the alternative theories of gravity (ATG) should  be in agreement with all previous successful tests/measurements which nominated general relativity as the prevailing theory of gravity. This means that ATGs should be in agreement  with general relativity in the, well tested, weak-field regime and the deviations should emerge when the field is strong. The gravitational field of neutron stars is quite strong and thus they can serve as candidates for testing ATGs.

Scalar-tensor theory of gravity is one of the most natural extensions of general relativity, where gravity in addition to the usual metric tensor $g_{\mu\nu}$ is mediated also by a massless long-range scalar field $\varphi$. The metric $g_{\mu\nu}$ describing the physical phenomena is connected to the metric in the Einstein frame $g_{*\mu\nu}$ via a conformal transformation of the form $g_{\mu\nu}=A^2(\varphi)g_{*\mu\nu}$. In this paper, we adopt the  conformal factor proposed by Damour and Esposito-Far\`{e}se \cite{DE1993}, \ie $A(\varphi)=\exp(\beta\varphi^2/2)$. Here $\beta$ is the coupling parameter, while for $\beta=0$ we recover  general relativity. The stellar equilibrium configurations in scalar-tensor gravity are  derived by integrating the modified Tolman-Oppenheimer-Volkoff (TOV) equations \cite{Harada1998,SK2004}, assuming a specific equation of state (EOS) and an asymptotical value of scalar field $\varphi_0$, here  we may be chosen to vanish, $\varphi_0=0$. For $\beta\lsim -4.35$ the structure of neutron star can dramatically deviate from the equivalent general relativistic models  \cite{Harada1998}, which is known as spontaneous scalarization. It is also found that the spontaneous scalarization for fast-rotating neutron stars, close to the Kepler limit, can set in for larger values of $\beta$ ($\beta\lsim -3.9$) \cite{DYSK2013}. This leads to even larger deviations in the mass vs radius diagram of the various EOSs and also significant deviations in the moment of inertia. For  the late-inspiral phase of binary neutron star systems, the possibility of the so-called dynamical scalarization has been also suggested. This can take place again for values of $\beta$ larger than $-4.35$ \cite{BPPL2013,PBPL2014,TSB2015}. Meanwhile, the observation of binary neutron star-white dwarf system set constraints in the lower value of $\beta$ \cite{Freire2012}, \ie they predict  in the most optimistic case  $\beta\gsim -5$.

In addition to the uncertainty for the correct  theory for gravity in the strong-field regime, the EOS for neutron star matter is also quite unconstraint \cite{shapiro-teukolsky}. In particular, constraining the EOS for high density region is still impossible via the terrestrial nuclear experiments due to the nature of nuclear saturation. So, the observation of neutron stars, in the electromagnetic and gravitational wave spectrum, via the associated phenomena (radio, X- and gamma rays, gravitational waves) are crucial for tracing the  EOS for the neutron star matter. Neutron star's mass is an important parameter which is typically the easiest to be estimated from the binary motion. Actually, the discovery of neutron stars with mass $\sim 2M_\odot$  \cite{D2010,A2013} led in the exclusion of some of the soft EOSs, which were predicting maximum mass less than $2M_\odot$.  While the recent detection of merging neutron stars led to some further constraints with regard to the softness/stiffness of the EOS as well as the radius \cite{Abbott2017}, see also \cite{Bauswein2017,Radice2017,Most2018}.

The compactness, which is the ratio of the mass over the radius, is a very important quantity in relativistic stars, and expresses the strength of the surface gravitational field.  As the compactness increases, the relativistic effects become more pronounced, and for example can affect the trajectories of photons emitted from the surface of neutron stars. In fact, the compactness would be determined via the observations of pulse profiles radiated from the neutron star by the operating x-ray timing mission with Neutron star Interior Composition ExploreR (NICER) \cite{NICER}, which may tell us the gravitational theory in the strong-field regime \cite{SM2017,S2017}. Moreover, the compactness is a key property for neutron star asteroseismology \cite{AK1998} becoming more pronounce for the mainly spacetime oscillations, the so-called $w$-modes \cite{KS1992}. To be more specific, the frequencies of $w$-mode gravitational waves from neutron stars can be well associated with the compactness \cite{AK1998}. In addition, if the compactness is too high and the stellar radius is smaller than the Regge-Wheeler potential barrier, whose position is a similar to the radius of the photosphere in the Schwarzschild spacetime, i.e., $r \sim 3M$, one may observe  a class  of spacetime perturbations, the so-called trapped $w$-modes \cite{CF1991,K1994}. In fact, the effective potential in the wave equation for spacetime oscillations for such  ultra-compact stars has a minimum inside the Regge-Wheeler potential barrier and this allows for (semi)bound states, the trapped $w$-modes.  This type of ultra-compact stars can be constructed in general relativity for some specific cases \cite{ultra-GR}, while the possibility of constructing such compact objects in scalar-tensor theory will be examined in this article.  Here we systematically examine effects of scalarization in the compactness of neutron stars.

The compactness of neutron stars in scalar-tensor gravity has already been discussed in Ref. \cite{TKW98}, where a modified Buchdahl inequality was derived and a theory-dependent maximum mass-to-size ratio was also expressed as a function of the Arnowitt-Deser-Misner (ADM) mass, scalar charge, position of stellar surface, and the coupling constant in gravitational theory. Then, by adopting the polytropic EOS, the maximum mass-to-size ratio was determined numerically. However, it will be interesting a study on the maximum allowed compactness for the more realistic case \ie by using data from terrestrial nuclear experiments and constraints imposed by observations in th electromagnetic and gravitational wave observations. In addition, since the Buchdahl limit in general relativity, which corresponds to $2M/R=8/9$, can only realize in the case of a uniform density star, it is better to examine such a case in scalar-tensor gravity, even though it may be more academic. Thus, in order to carefully examine the maximum compactness of a scalarized neutron star, we adopt a relatively simple EOS characterized by the nuclear saturation parameter in the lower density region and the sound velocity in higher density region. In order to explore the existence of ultra-compact stars, we also numerically study the Tolman VII solution in scalar-tensor gravity. The Tolman VII solution is an analytical solution in general relativity \cite{TolmanVII}, with a free scaling parameter which can mimic very well the overall properties of realistic EOS and in the limit one can get even the uniform density solution. Its stability was re-examined recently in \cite{Moust2017}. This attractive feature of Tolman VII solution led us in using it for examining the neutron star compactness in scalar-tensor gravity. In this paper, we adopt geometric units, $c=G_*=1$, where $c$ and $G_*$ denote the speed of light and the gravitational constant, respectively, and the metric signature is $(-,+,+,+)$.

\section{Compactness for the maximum mass neutron star}
\label{sec:II}

The compactness ($M/R$) gets its maximum value for maximum mass allowed by a specific EOS. Thus the search for the  maximum compactness is equivalent to the search for the maximum mass for a specific EOS.  Meanwhile, the dynamical stability {criterion} for the neutron stars in scalar-tensor gravity is not yet properly studied. Here we  assume, by following the general relativistic methodology, that the scalarized neutron star  becomes dynamically unstable if $\partial M_{\rm ADM}/\partial R >0$ in the $M_{\rm ADM}-R$ plane and in addition the central density is larger than {that of the scalarized neutron star model with maximum mass}. Here $M_{\rm ADM}$ denotes the ADM mass, which coincides with the gravitational mass in general relativity. The maximum mass limit for neutron stars has been already discussed in general relativity (\eg \cite{H1978,LPMY1990,KB1996,KSF1997,Sotani2016}) and recently in scalar-tensor gravity \cite{SK2017}. In the later study, two assumptions were set for the approximate  EOS used, \ie  first that the sound velocity should not exceed the speed of light and second that the EOS describing the core region should be smoothly connected to the EOS describing the exterior low density region which is constrained by the nuclear theory and experiments. More specifically, in order to group the many uncertainties associated with the EOS in the high density region, we adopted a simple analytic description of the EOS \cite{Sotani2016,SK2017}, characterized by the maximum sound velocity in a high density region, $v_s$. This EOS can be written in the form 
\begin{equation}
  p = p_t + v_s^2 \left(\varepsilon - \varepsilon_t\right),  \label{eq:eos1}
\end{equation}
where the pressure, $p$, is expressed as a function of the energy density, $\varepsilon$, in the high density region ($\varepsilon\ge\varepsilon_t$) and it is connected to the EOS in the low density region at $\varepsilon=\varepsilon_t$, with the transition pressure given by $p=p_t$. Actually, the value of $p_t$ is determined by the EOS used in the low density region with $\varepsilon=\varepsilon_t$.

Unlike in the high density region, the EOS in the low density region could be constrained via the terrestrial nuclear experiments. In practice, the bulk energy per baryon of uniform nuclear matter at zero temperature for any EOSs can be expanded in the vicinity of the saturation point as a function of the baryon number density and the neutron excess \cite{L1981}. The coefficients in such an expansion are known as nuclear saturation parameters, and are constrained via the terrestrial nuclear experiments (e.g., \cite{LL2013}). Even so, the constraint on the incompressibility of symmetric nuclear matter, $K_0$, and the density dependent nuclear symmetry energy, $L$, are relatively weaker constraint than the other saturation parameters. The current constraints on $K_0$ and $L$ are estimated to be $K_0=230\pm40$ MeV \cite{Khan2013} and $30\lsim L\lsim 80$ MeV \cite{Newton2014} by numerous terrestrial nuclear experiments. Moreover, the nuclear saturation parameters  play an important role in the description of the low mass neutron stars. In fact, the neutron star models with central density less than twice the saturation density, can be well described by  the combination of $K_0$ and $L$ via $\eta\equiv(K_0L^2)^{1/3}$ \cite{SIOO2014,SSB2016}. The current constraints on $K_0$ and $L$ lead to the plausible range of $\eta$ that is $55.5\lsim \eta\lsim 120$ MeV. Thus, the EOS in high density region described by Eq. (\ref{eq:eos1}) is matched to the lower density region EOS characterized by $\eta$ at the transition density $\varepsilon_t$, which is set to be twice the saturation density. Here, for the low density region, we adopt the phenomenological EOS derived by Oyamatsu and Iida \cite{OI2003,OI2007}. This EOS is constructed for optimizing the nuclear saturation parameters except for $K_0$ and $L$ in such a way as to obey the experimental data, given values of $K_0$ and $L$. This choice leads to an EOS (for the low and high density regime) which is characterized by only two parameters, i.e., $v_s$ and $\eta$.

\begin{figure}
\begin{center}
\includegraphics[scale=0.5]{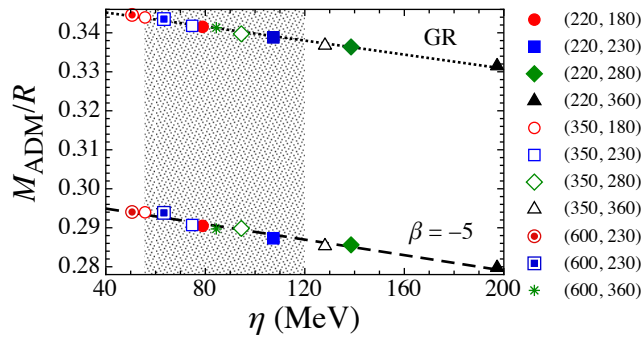} 
\end{center}
\caption{
The stellar compactness for the maximum mass neutron star in general relativity and in scalar-tensor gravity ($\beta=-5$) is shown as a function of the nuclear saturation parameter $\eta$. The different marks correspond to the different equation of state characterized by the nuclear saturation parameters $(y, K_0)$. The shaded region denotes the plausible range of value of $\eta$, which is constrained via the terrestrial nuclear experiments. 
}
\label{fig:MR-eta}
\end{figure}

In this paper, since we especially focus on the maximum possible mass of neutron stars,  we put $v_s=1$ in Eq. (\ref{eq:eos1}). As in Refs. \cite{Sotani2016,SK2017}, we construct the maximum mass neutron star models in general relativity and the scalarized models with maximum mass for $\beta=-5$. The mass of scalarized neutron stars for the values of $\beta$ allowed by observations can not reach the maximum mass of the corresponding models in general relativity for the case of $v_s^2\gsim 0.6$ \cite{SK2017}. The maximum compactness in general relativity and in scalar-tensor gravity with various values of $\eta$ is plotted in Fig. \ref{fig:MR-eta}.  In this figure, the different marks correspond to  different EOS models characterized by the saturation parameters $y\equiv -K_0S_0/(3n_0L)$ and $K_0$, where $n_0$ and $S_0$ respectively denote the saturation density of symmetric nuclear matter and the nuclear symmetry energy at the saturation point. For reference, the plausible values of $\eta$ are also shown by the shaded region. From this figure, it becomes obvious that as for the case of the maximum mass in general relativity and in scalar-tensor gravity, the maximum compactness is well fitted as a linear function of $\eta$, having the form
\begin{equation}
   \frac{M_{\rm ADM}}{R} = a_1 - a_2 \times \left(\frac{\eta}{1\ {\rm MeV}}\right) \, .
\end{equation}
Here the coefficients $a_1$ and $a_2$ depend on the coupling parameter $\beta$, i.e., $a_1=0.3488$ and $a_2=8.9952 \times 10^{-5}$ in general relativity ($\beta=0$), while $a_1=0.2989$ and $a_2=9.9510 \times 10^{-5}$ in scalar-tensor gravity with $\beta=-5$. The fitting lines are also plotted in Fig. \ref{fig:MR-eta}, where the dotted line corresponds  to general relativistic case and the dashed line to scalar-tensor gravity with $\beta=-5$. Thus, by adopting the plausible values of $\eta$, we find that the maximum compactness reaches $M_{\rm ADM}/R=0.344 - 0.338$ for $\eta=55.5 - 120$ MeV in general relativity, while $M_{\rm ADM}/R=0.293 - 0.287$ for $\eta=55.5 - 120$ MeV in scalar-tensor gravity with $\beta=-5$. In all cases, the maximum compactness of the scalarized neutron stars is smaller than that in general relativity.

\section{Tolman VII solutions}
\label{sec:III}

Now, we examine the Tolman VII solutions in scalar-tensor gravity. Tolman VII solution is known as an analytic solution in general relativity \cite{TolmanVII}, where the energy density distribution inside the star with a given radius, $R=R_b$, is provided by the relation
\begin{equation}
  \varepsilon = \varepsilon_c\left[1-(1-w)\left(\frac{r}{R_b}\right)^2\right].   \label{eq:er}
\end{equation}
Here, $\varepsilon_c$ denotes the central energy density and $w$ is the ratio of the energy density at the stellar surface to $\varepsilon_c$. In particular, the case with $w=1$ corresponds to a uniform density star. The quadratic falloff in the density is a very good approximation to most of the EOS \cite{LP2001}. Typically the differences of the various EOSs are minor if one compares only the density profiles. For this reason, this exact solution of the Einstein field equations provides a unique tool for testing the fundamental properties of neutron stars.

The density profile shown in Eq. (\ref{eq:er}), allows for the analytic derivation, in general relativity, of the stellar mass
\begin{equation}
  M=\frac{4\pi}{15}(2+3w)\varepsilon_cR^3.   \label{eq:mass}
\end{equation}
In scalar-tensor theory, if  one adopts the Tolman VII density profile given by Eq. (\ref{eq:er}), there is no obvious way to get an analytic solution and thus for constructing equilibrium configurations one needs to integrate numerically the associated TOV equations \eg \cite{Harada1998, SK2004}. In the numerical integration of the TOV equations the radius is be defined via the vanishing of the pressure at $r=R_b$. Since $R_b$ is given in Eq. (\ref{eq:er}), one should numerically find the central pressure so that the pressure vanishes at $r=R_b$ for given values of $\varepsilon_c$ and $R_b$. Moreover, the TOV equations in scalar-tensor gravity are written and solved in the so-called Einstein frame, thus one has to covert the stellar parameters into the physical (Jordan) frame, i.e., the stellar radius in the Einstein frame, $r=R_b$, should be converted to the circumference radius, i.e., $R=R_b\exp(\beta\varphi_s^2/2)$, where $\varphi_s$ denotes the surface value of scalar field.

\begin{figure*}
\begin{center}
\begin{tabular}{cc}
\includegraphics[scale=0.5]{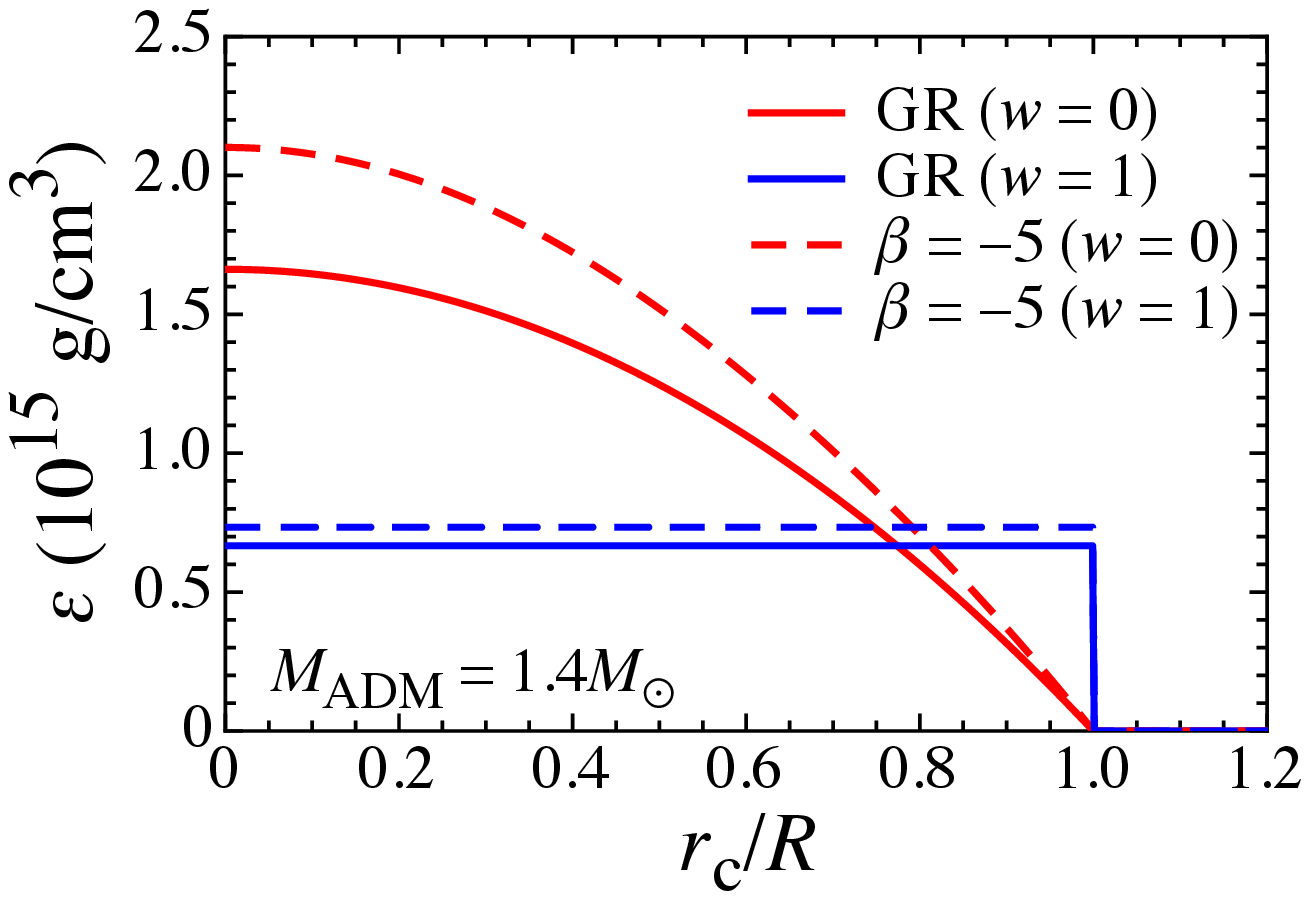} &
\includegraphics[scale=0.5]{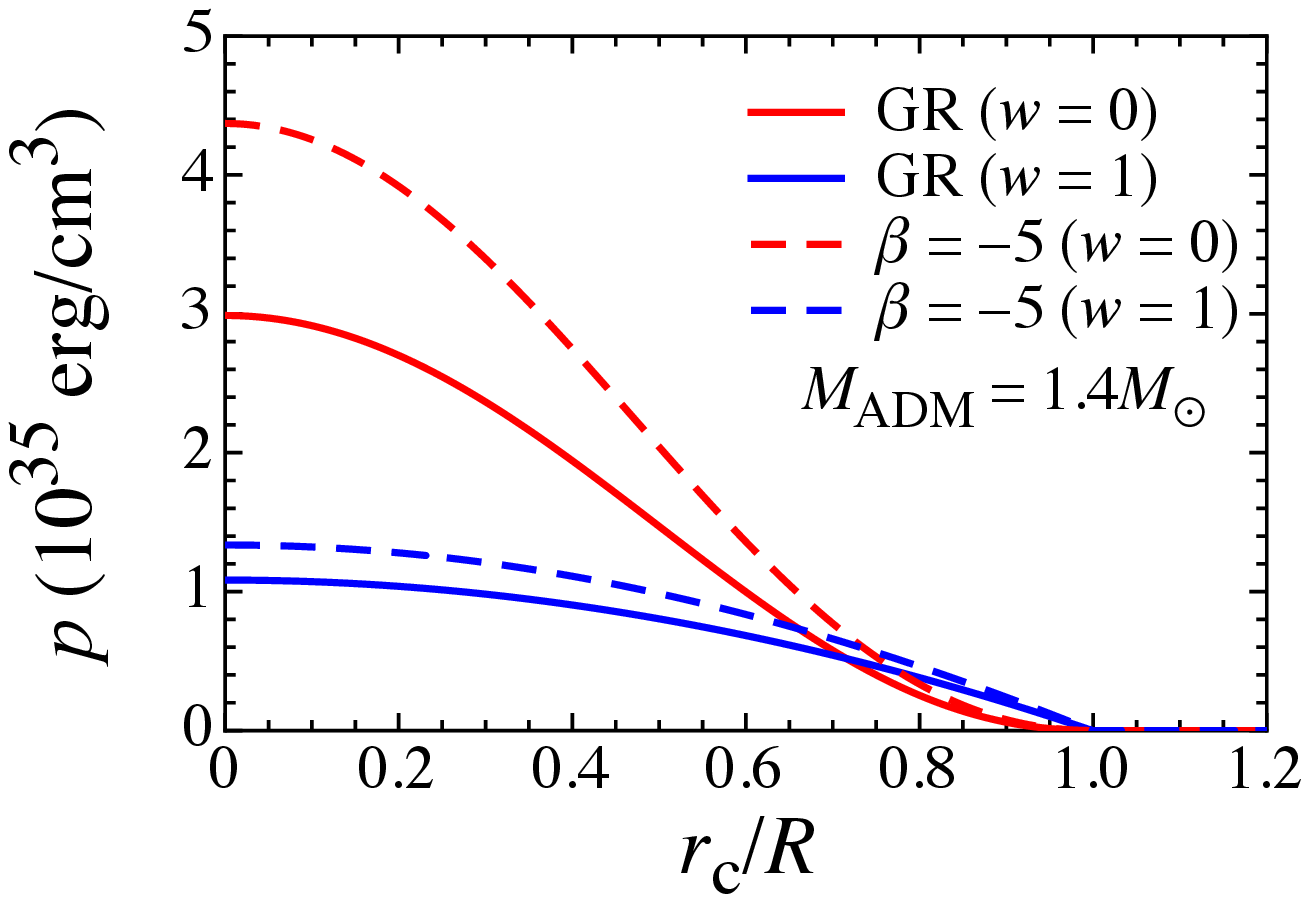} \\
\includegraphics[scale=0.5]{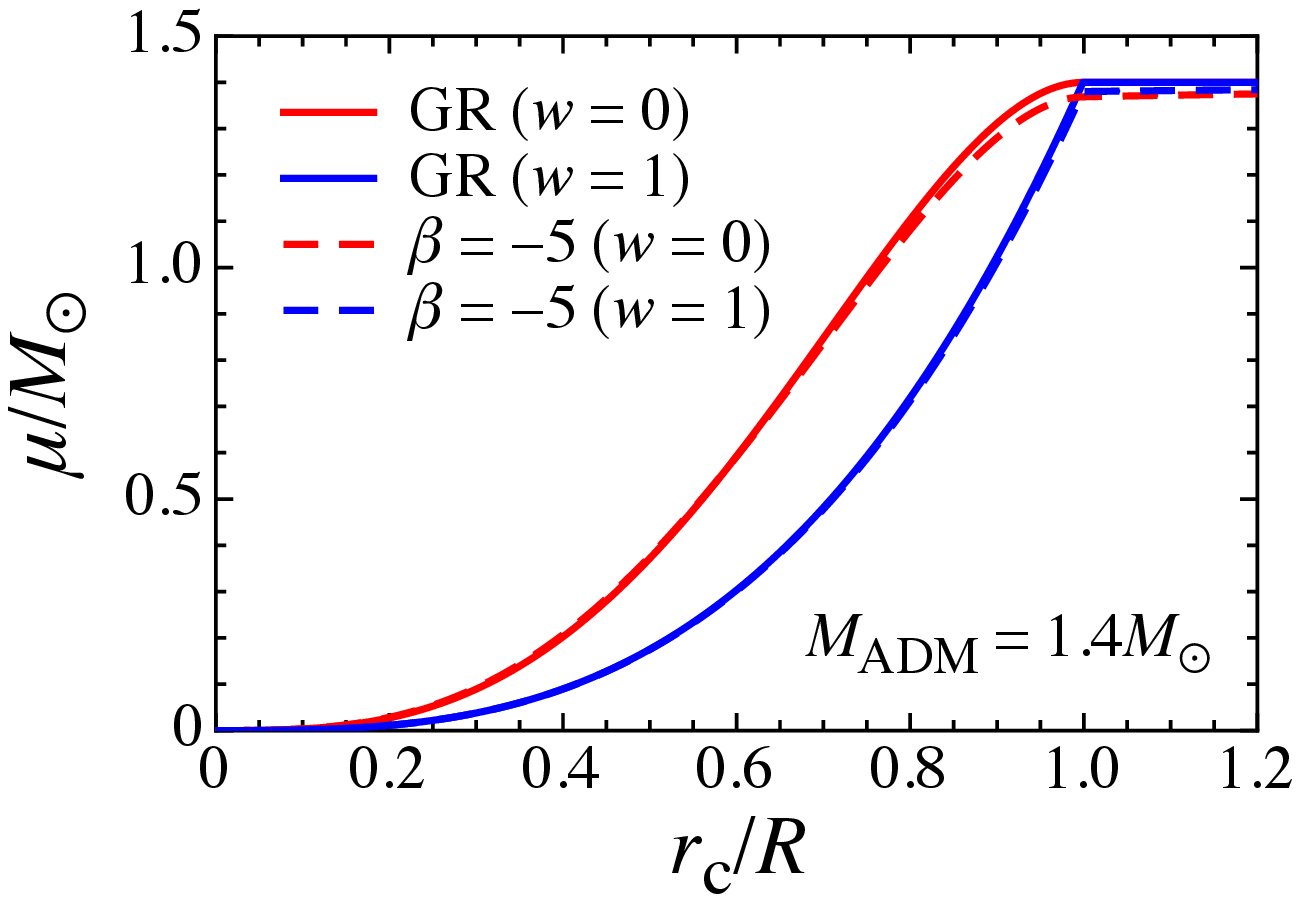} &
\includegraphics[scale=0.5]{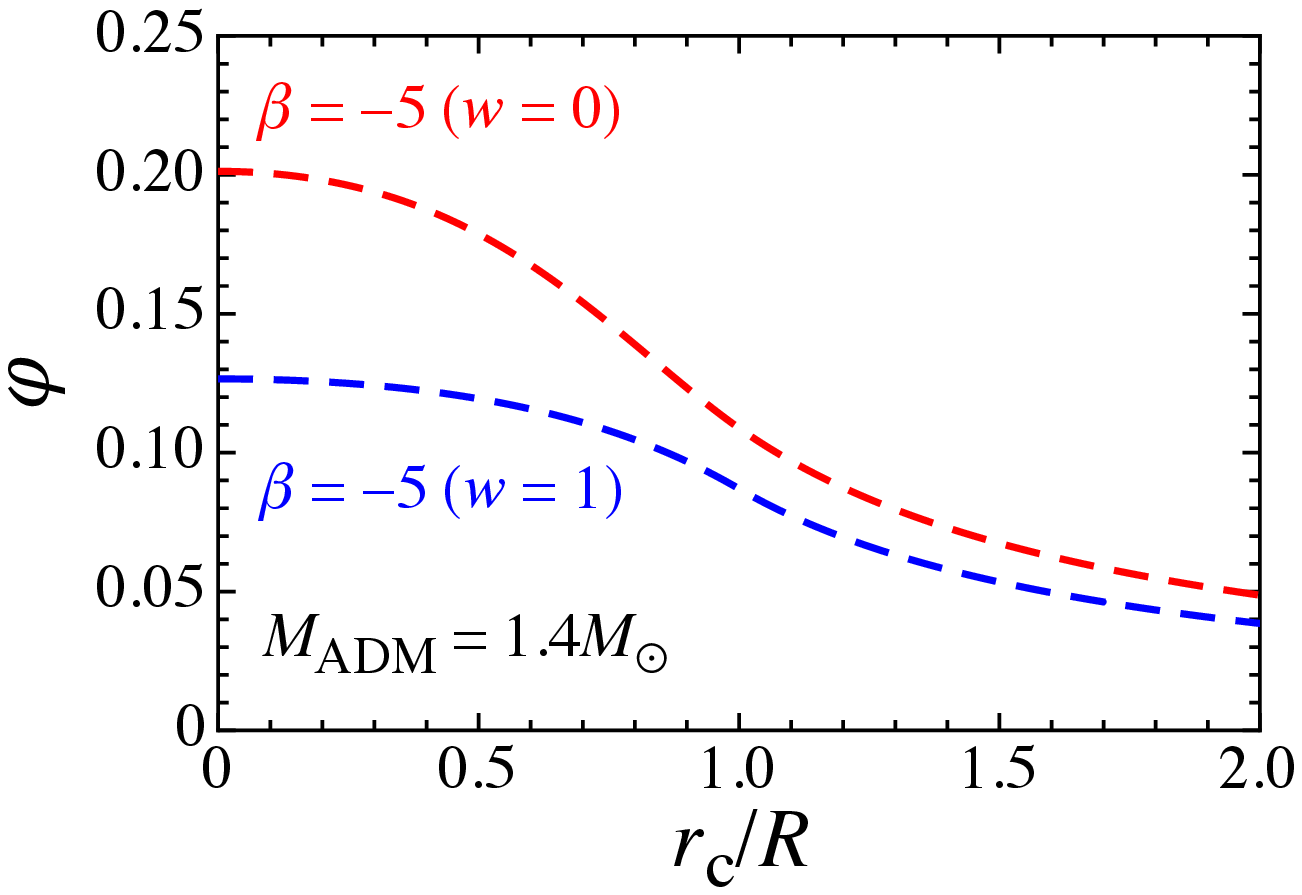} 
\end{tabular}
\end{center}
\caption{
The radial profiles of energy density $\varepsilon$, pressure $p$, mass function $\mu$, and scalar field $\varphi$ for the Tolman VII solutions with $w=0$ and 1 in general relativity and in scalar-tensor gravity with $\beta=-5$ are plotted. Here the stellar mass is fixed to  $M_{\rm ADM}=1.4M_\odot$ and $r_c$ denotes the circumference radius determined by $r_c=r\exp(\beta\varphi^2/2)$.
}
\label{fig:Rb10M14}
\end{figure*}

In Fig. \ref{fig:Rb10M14}, for instance, the radial profiles of the energy density $\varepsilon$, pressure $p$, mass function $\mu$, and scalar field $\varphi$ for a specific stellar model with $M_{\rm ADM}=1.4M_\odot$ are shown as a function of $r_c/R$ in general relativity and in scalar-tensor gravity with $\beta=-5$ for the case of $w=0$ and 1, where $r_c$ denotes the circumference radius determined by $r_c=r\exp(\beta\varphi^2/2)$. From this figure, one can observe that the profile of $\mu$ is almost independent of the chosen theory of gravity for fixed values of $w$, while the other quantities ($\varepsilon$, $p$) depend stronger on the chosen theory.

\begin{figure*}
\begin{center}
\begin{tabular}{ccc}
\includegraphics[scale=0.43]{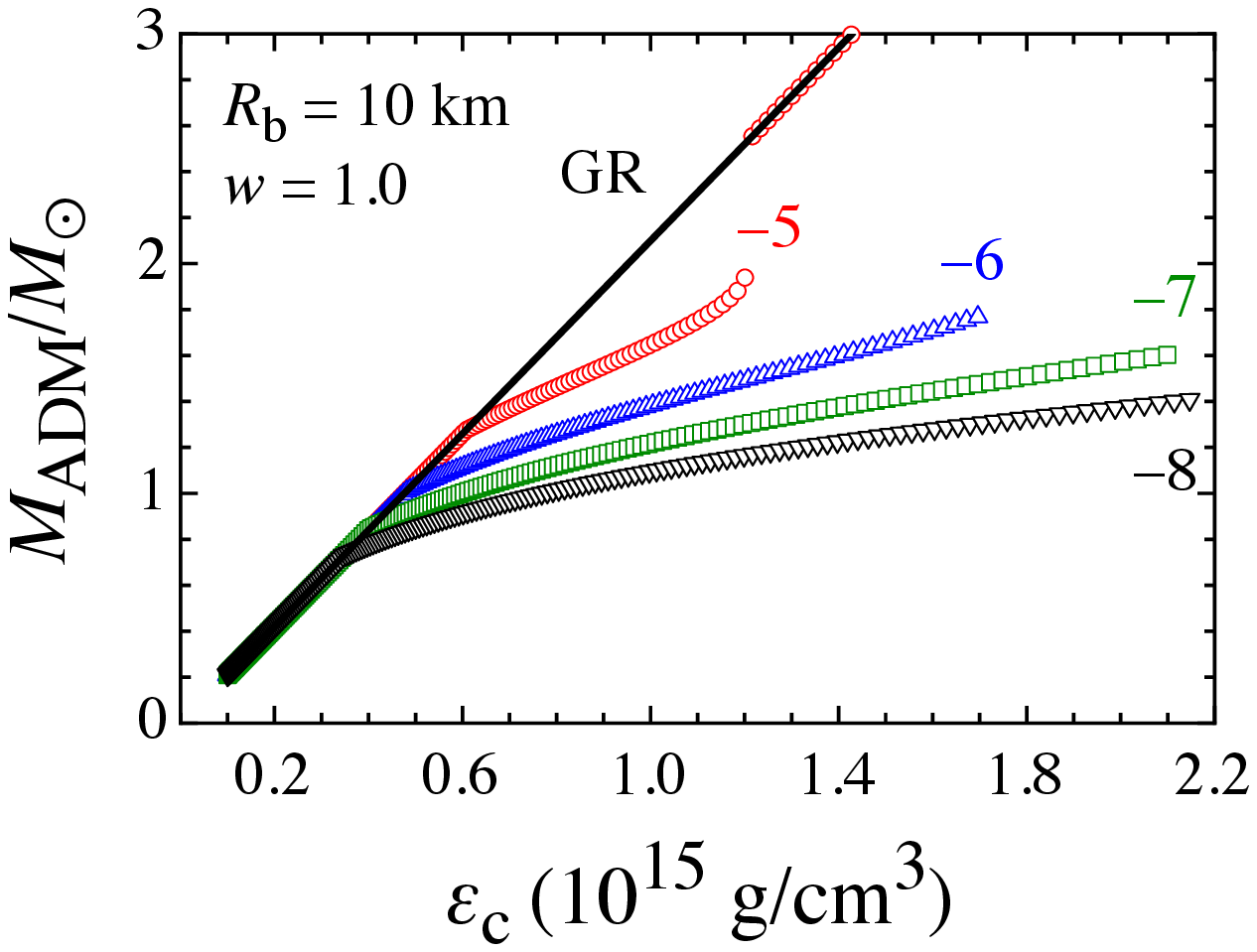} &
\includegraphics[scale=0.43]{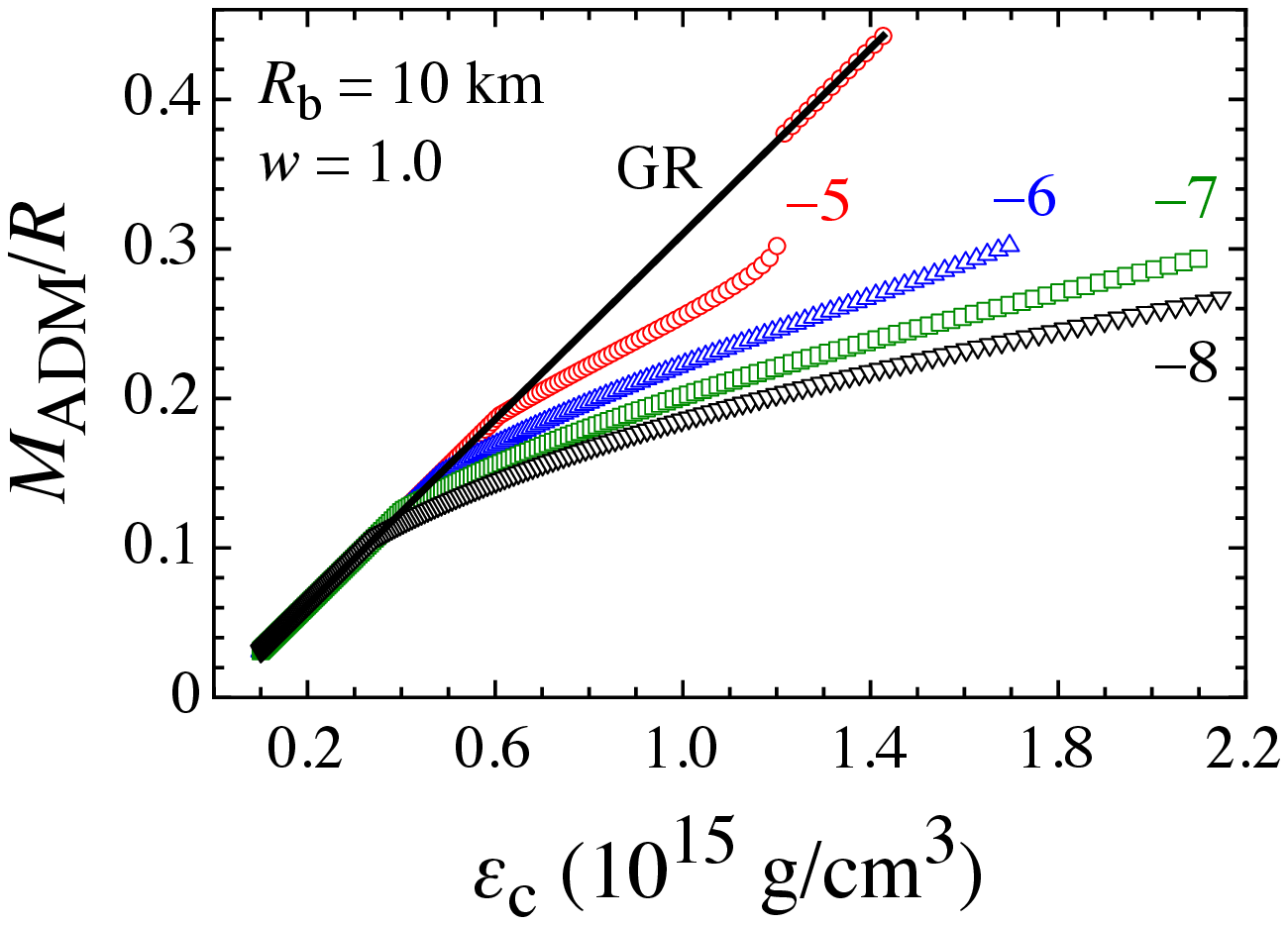} &
\includegraphics[scale=0.43]{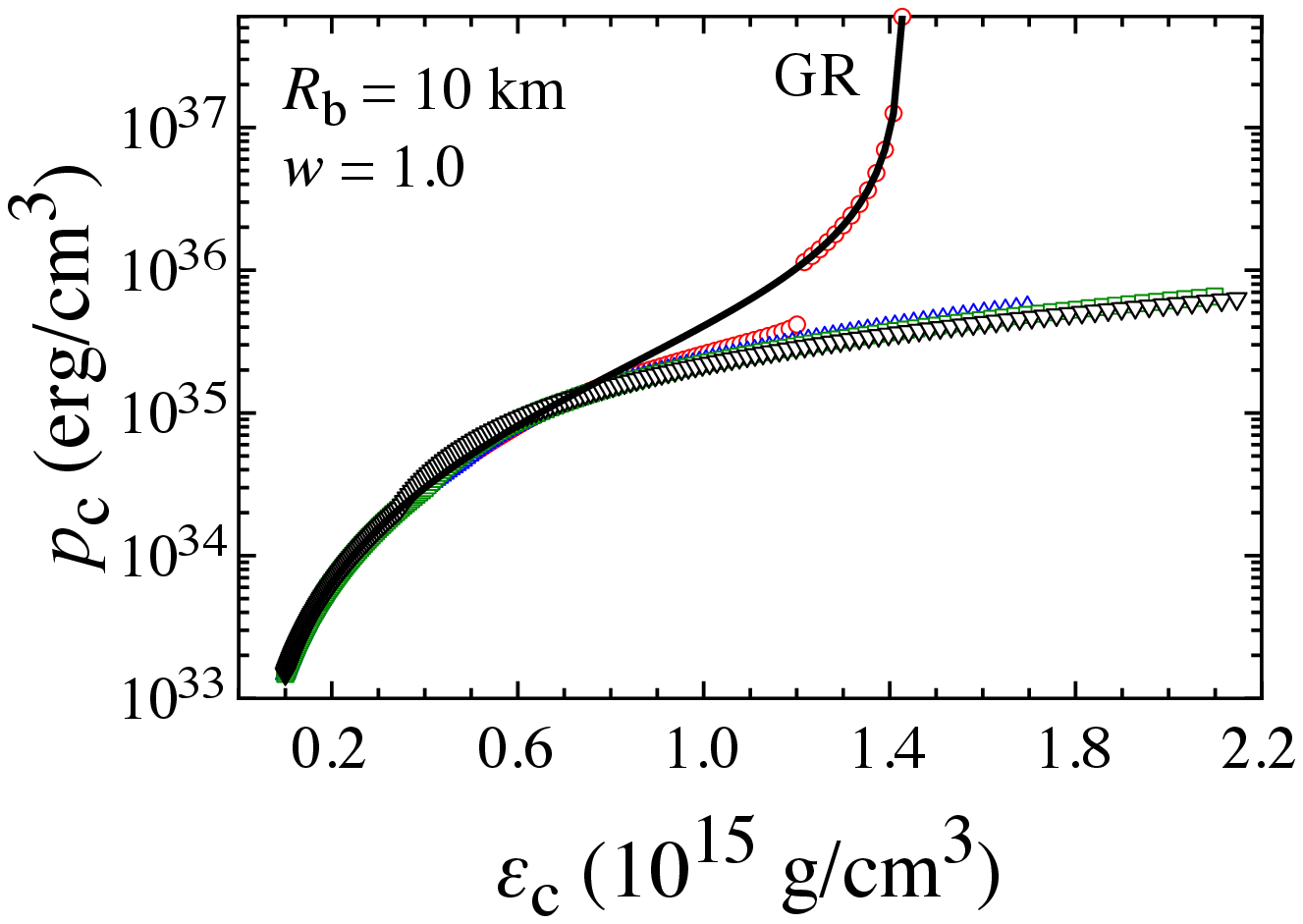} 
\end{tabular}
\end{center}
\caption{
The ADM mass (left panel), the compactness (middle panel), and the central pressure (right panel) for different values of the coupling constant $\beta$ are plotted as function of $\varepsilon_c$ (central energy density) for the case of $R_b=10$ km and $w=1$. The solid lines are the results derived in general relativity, while the others are derived in scalar-tensor gravity.
}
\label{fig:Rb10w10a}
\end{figure*}

The ADM mass, the compactness and the central pressure for sequences of neutron stars models based on the Tolman VII solution with $R_b=10$ km and $w=1$ are shown in Fig. \ref{fig:Rb10w10a} as a function of the central energy density. Here, we adopted four values for $\beta$, which are not actually  supported by observations. This choice was made in order to make more transparent the dependence of the equilibrium quantities on the scalar field. 

\begin{figure*}
\begin{center}
\begin{tabular}{cc}
\includegraphics[scale=0.5]{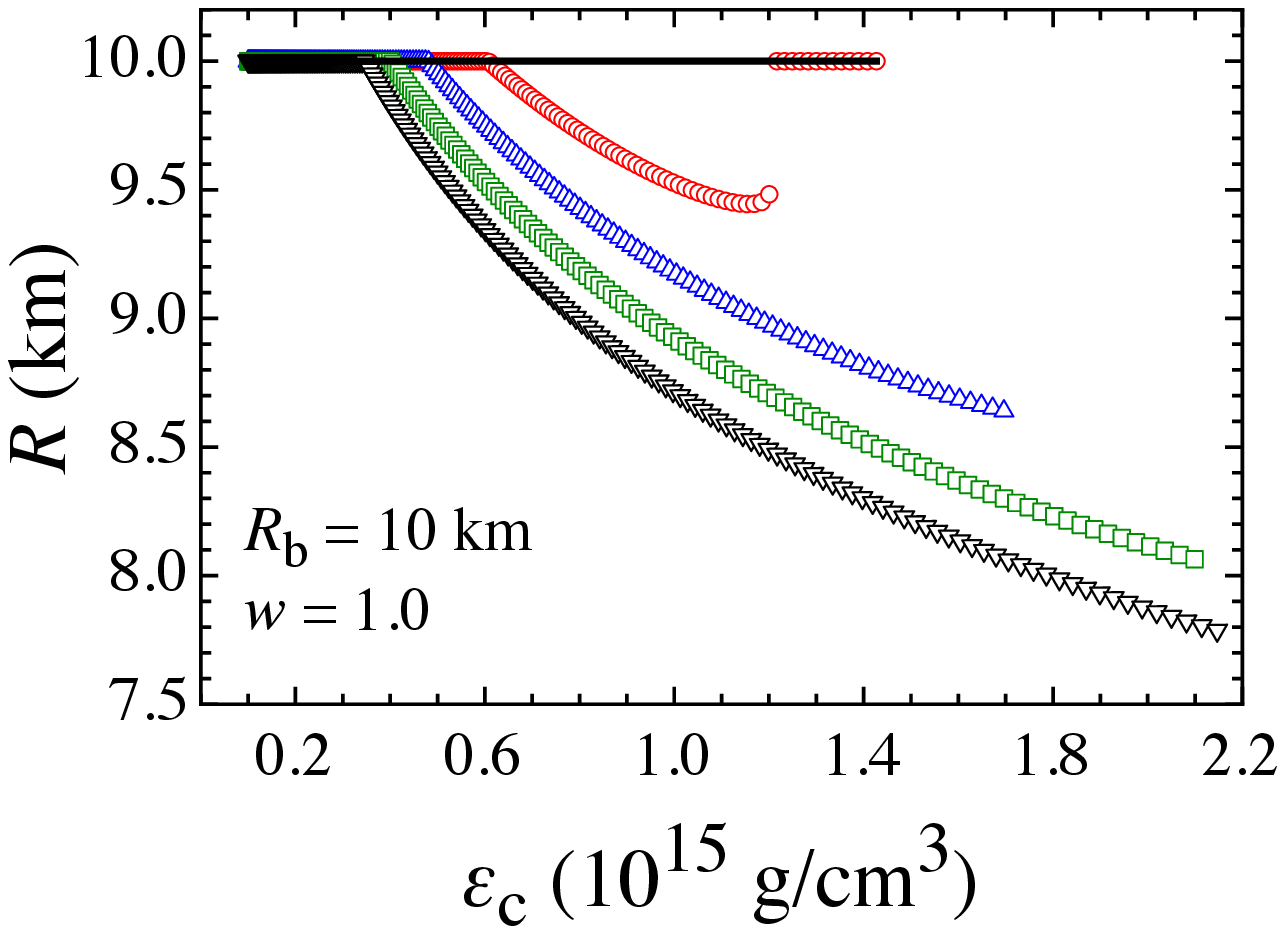} &
\includegraphics[scale=0.5]{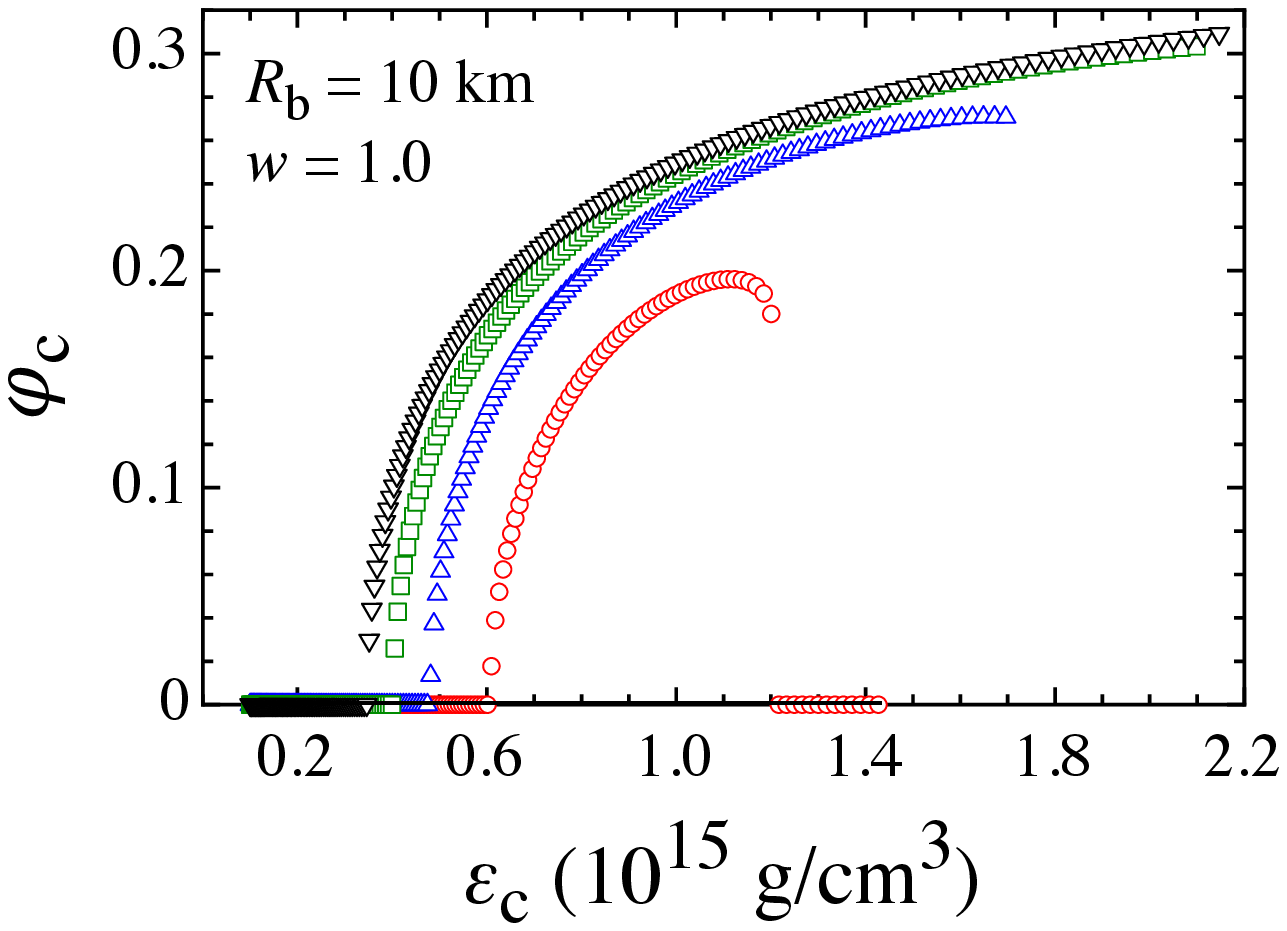} 
\end{tabular}
\end{center}
\caption{
The radius (left panel) and the central value of scalar field (right) panel with various values of coupling parameter $\beta$ are shown as a function of $\varepsilon_c$ for the case of $R_b=10$ km and $w=1.0$. The different lines are parametrized as in Fig. \ref{fig:Rb10w10a}.
}
\label{fig:Rb10w10b}
\end{figure*}

From these numerical calculations, we find that they reach the maximum allowed values of central density for $\beta$ = $-6$, $-7$ and $-8$,  while for smaller values \ie for $\beta=-5$ when the central energy density exceeds a critical value the stellar models in scalar-tensor gravity   ``jump'' into those of general relativity.  Hence, in this case the stellar models in scalar-tensor gravity, coincide with the general relativistic models.
We remark that for stellar models constructed for polytropes and realistic EOSs in scalar-tensor gravity the solutions smoothly merge with the general relativistic ones \cite{SK2004}.  The jumps observed in Figs. \ref{fig:Rb10w10a} and \ref{fig:Rb10w10b} seem to be a peculiarity of the EOS based on the Tolman VII solution because one needs to fix the radius in the Einstein frame $R_b$ in advance.

In this figure we also show that the mass and compactness of the scalarized Tolman VII solutions are always smaller than those expected in general relativity. Moreover, the stellar mass of scalarized Tolman VII models with  fixed central energy density decreases as $-\beta$ increases. In addition, we can see the existence of the critical (maximum) value of central energy density for which scalarization is possible. This critical value increases as $-\beta$ increases. An equivalent critical value for the central energy density ($\bar{\varepsilon}_c$) exists also in general relativity when the central pressure diverges. Technically, it seems to be possible to construct the scalarized Tolman VII solutions with the central energy density larger than $\bar{\varepsilon}_c$, for large values of  $-\beta$. We also observe that the central energy density, for which the scalarization sets in, decreases as $-\beta$ increases. For the same set of the Tolman VII parameters ($R_b=10$ km and $w=1$) used in Fig. \ref{fig:Rb10w10a}, we plot in Fig. \ref{fig:Rb10w10b} the stellar radius and the central value of scalar field   as functions of the central energy density. For increasing  $-\beta$ the scalar field's  contribution also increases while the radius of scalarized Tolman VII solution decreases even for fixed values of $R_b$.

\begin{figure*}
\begin{center}
\begin{tabular}{cc}
\includegraphics[scale=0.5]{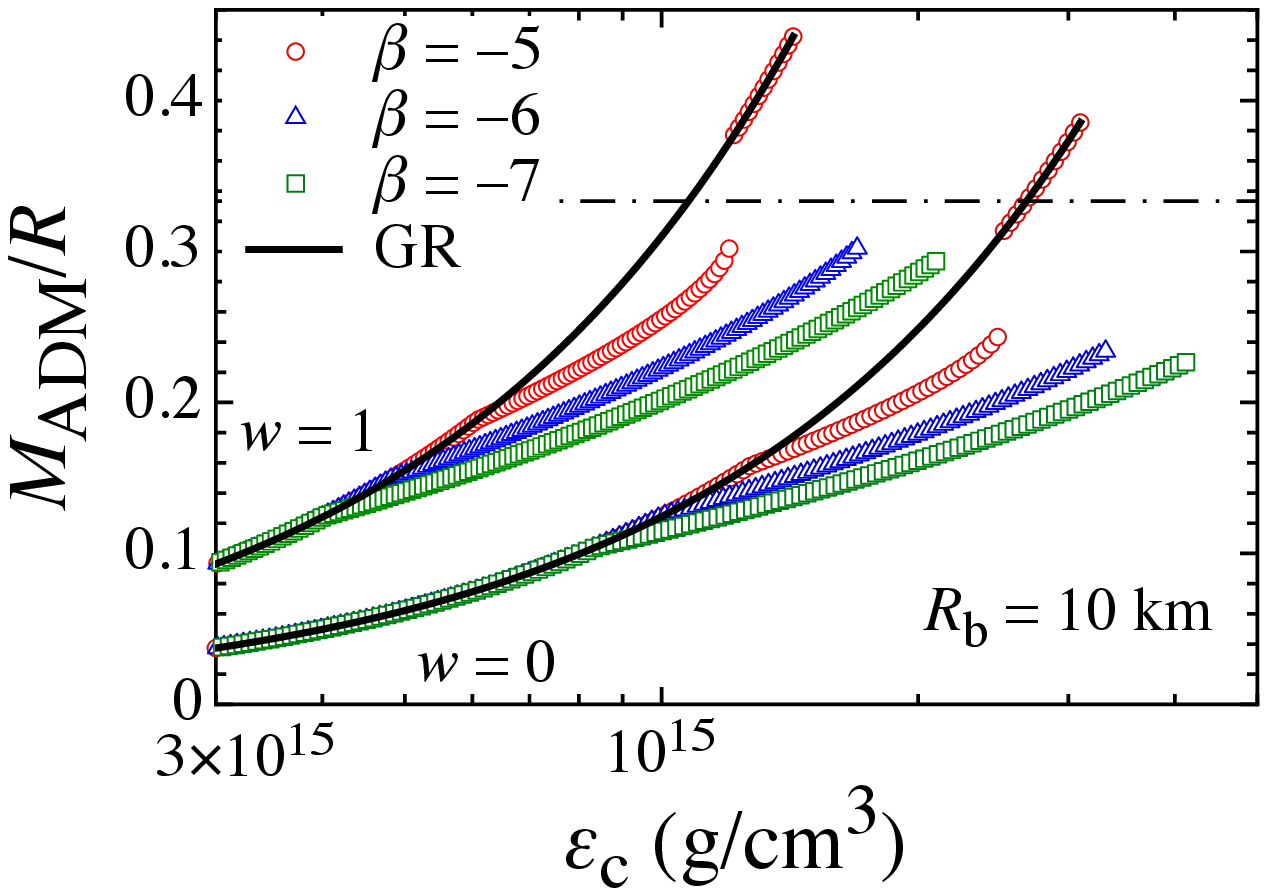} &
\includegraphics[scale=0.5]{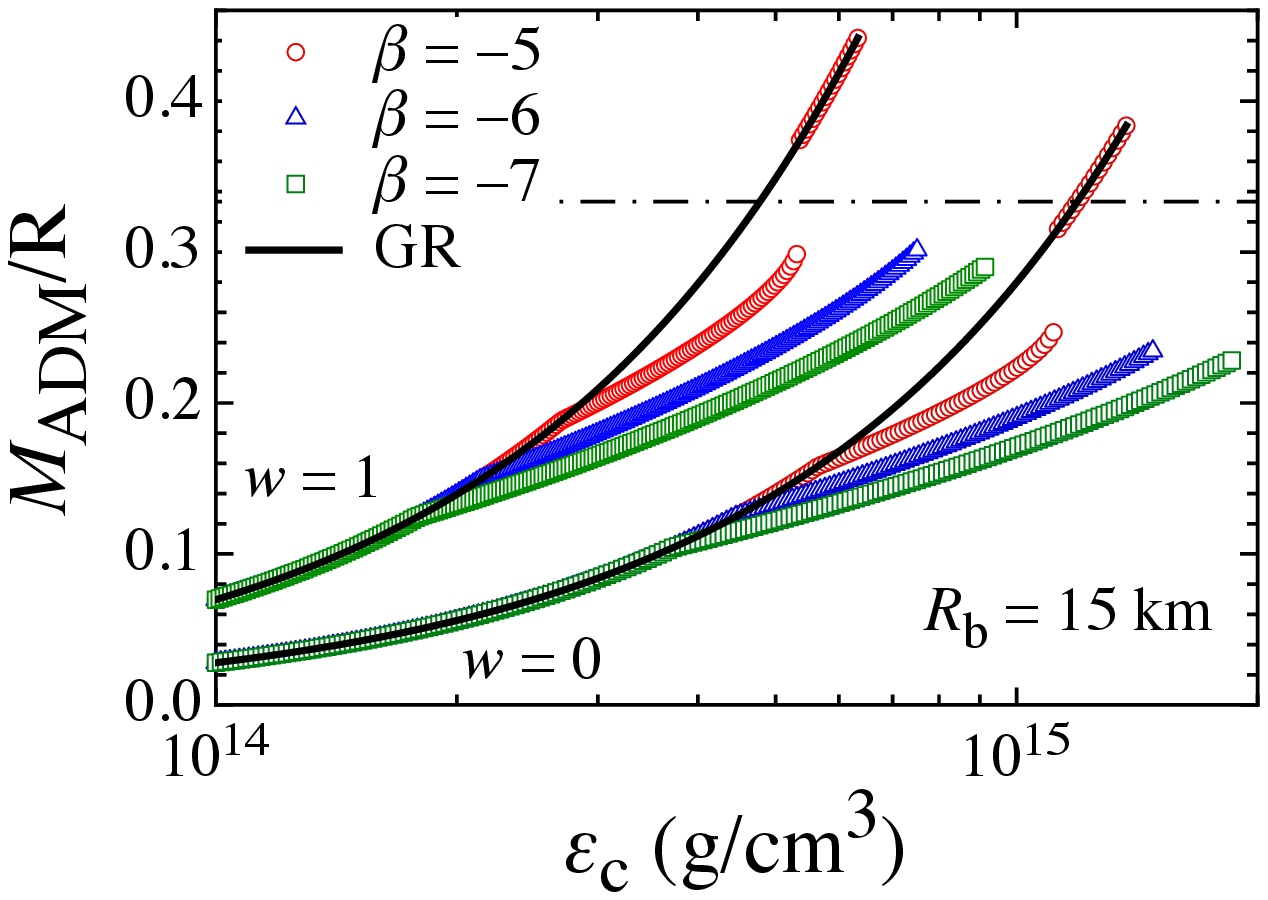} 
\end{tabular}
\end{center}
\caption{
Stellar compactness of the Tolman VII solutions with $w=0$ and 1 in scalar-tensor gravity with various coupling constant $\beta$ , where the left and right panels correspond to the results with $R_b=10$ and 15 km. The horizontal dot-dash-line denotes the radius of photosphere of the Schwarzschild spacetime, i.e., $R=3M_{\rm ADM}$.
}
\label{fig:Rb1015}
\end{figure*}

The neutron star compactness affects both the electromagnetic (\eg x-ray emission) and gravitational wave spectra and it is among the first parameters to be constrained by observations. Thus one may try to seek the effect of scalarization on the compactness. Here, we studied the effect of scalarization for a combination of values for $w$ and $R_b$ and the results are presented 
in Fig. \ref{fig:Rb1015}.
The compactness of the models based on the Tolman VII solution are plotted as function of the central energy density for $w=0$ and $w=1$. 
In the two panels we present the results for models with  $R_b=10$ and 15 km. From this figure, one can see that the compactness of the scalarized models is always smaller than the corresponding ones in general relativity for the same value of  central energy density. In addition, the compactness of the scalarized Tolman VII can not be higher than $M_{\rm ADM}/R=1/3$, which corresponds to the radius of the  photosphere of the Schwarzschild spacetime, as it can be observed from Fig. \ref{fig:Rb1015}.

\begin{figure*}
\begin{center}
\begin{tabular}{cc}
\includegraphics[scale=0.5]{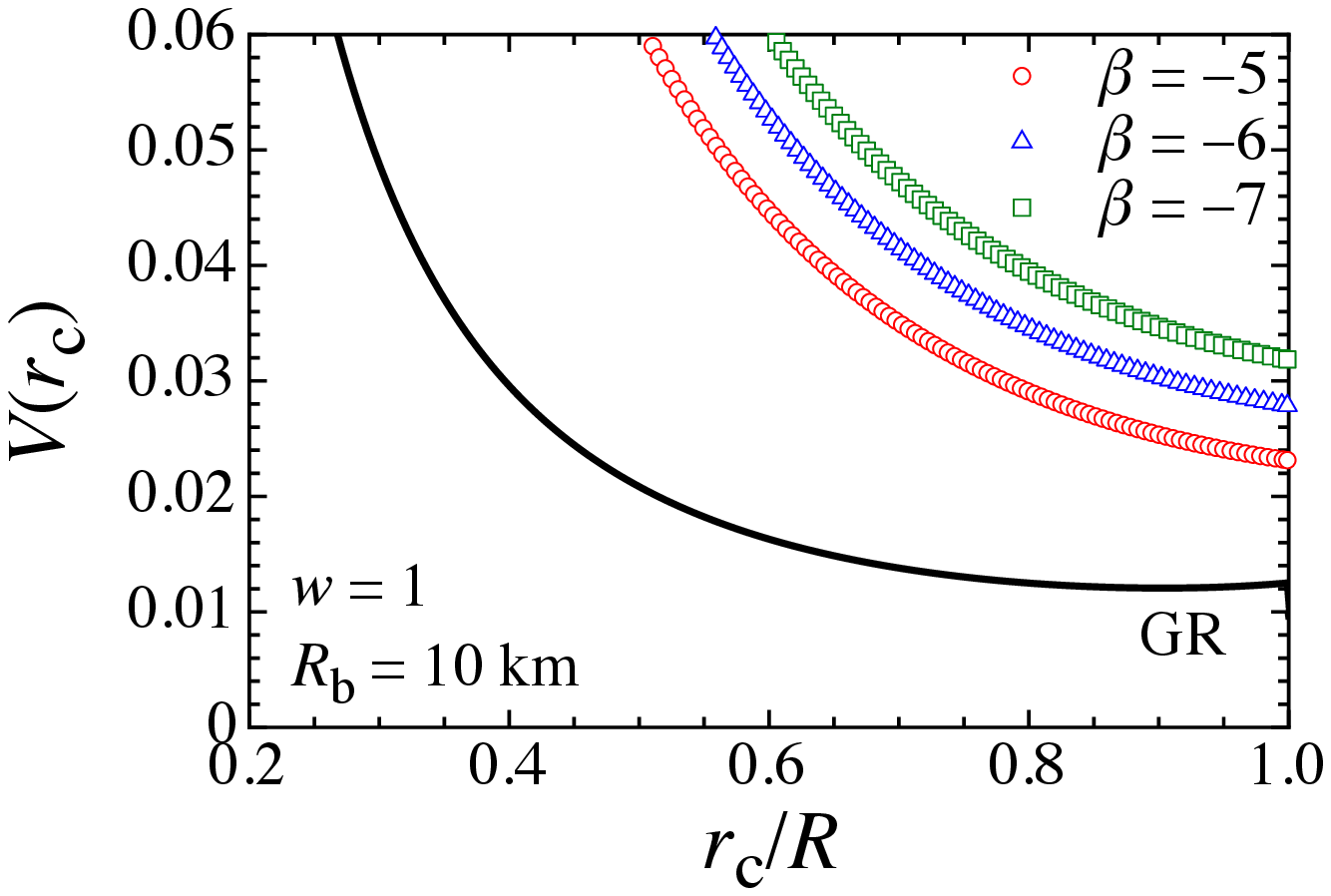} &
\includegraphics[scale=0.5]{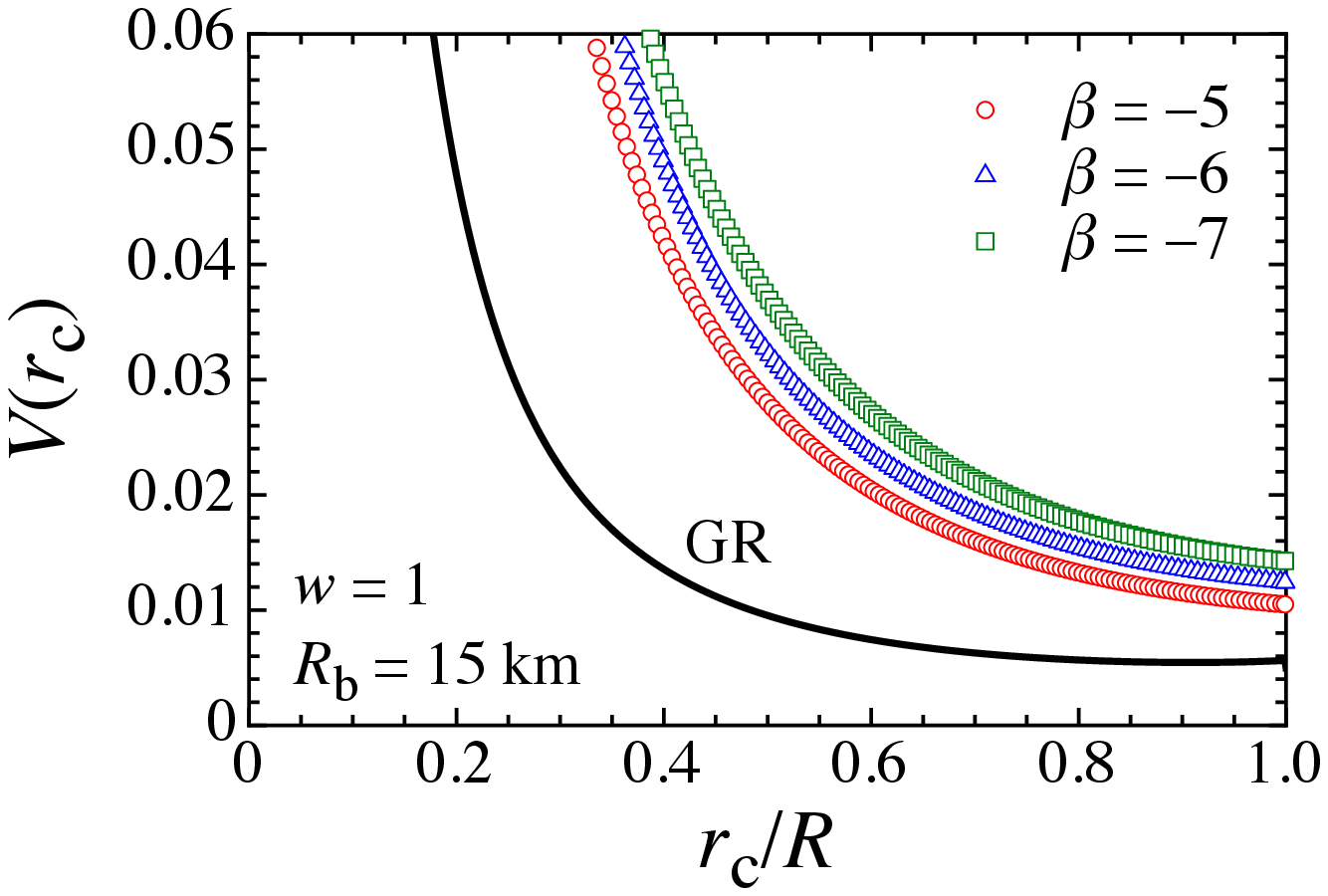} 
\end{tabular}
\end{center}
\caption{
Behavior of the effective potential of the axial $w$-mode gravitational oscillations is shown for the scalarized stellar models with maximum compactness, where $w$ is set to be 1 and $R_b$ is 10 km (left panel) and $R_b=15$ km (right panel) for various values of $\beta$. For reference the result in general relativity is also shown for the stellar model constructed with the same central energy density as the model with $\beta=-5$.
}
\label{fig:Vr}
\end{figure*}

The results, up to now suggest that  one cannot construct ultra-compact scalarized Tolman VII solutions, but this is still uncertain because the radius of the photosphere in scalar-tensor gravity may not be the same as that in the Schwarzschild spacetime. To confirm whether ultra-compact scalarized stellar models can be constructed or not, we check the behavior of the effective potential of the axial $w$-mode gravitational oscillations. In fact, the potential has a minimum inside the star if a stellar model is ultra-compact star. The effective potential for $\ell$-order oscillations is expressed as
\begin{equation}
  V = e^{2\Phi}\left[\frac{\ell(\ell+1)}{r^2} - \frac{6\mu}{r^3} + 4\pi G_*\left(\varepsilon - p\right)A^4\right],
\end{equation}
where $e^{2\Phi}$ corresponds to the $(t,t)$-component of metric in the Einstein frame \cite{SK2004}, and $G_*$ is the bare gravitational coupling constant. Finally, the mass function $\mu$ is defined by the $(r,r)$-component, $g_{*rr} = (1-2\mu/r)^{-1}$, of the metric in the Einstein frame and is directly connected to the ADM mass via Eq.  (31) in Ref. \cite{SK2004}. In Fig. \ref{fig:Vr}, we plot the effective potential with $\ell=2$, for the scalarized stellar models with maximum compactness for $w=1$, where the left and right panels correspond to the results for $R_b=10$ km and 15 km. In the same figure, for reference we show the effective potential in general relativity for the stellar model constructed with the same central energy density as the model with $\beta=-5$. From this figure, one can see that the effective potential in scalar-tensor gravity monotonically decreases even for the scalarized stellar model with maximum compactness, while the equivalent general relativistic model has a minimum inside the star although for the specific model it is very faint since the compactness is not so high. 
This means one cannot construct ultra-compact scalarized Tolman VII solutions, which can be the basis for trapped $w$-modes \cite{CF1991,K1994}, the ergoregion instabilities in the presence of rotation \cite{KRA2004,Card2008} and not even to serve as exotic echo producing objects \cite{Abedi2017,CP2017,MVK2017}.

\begin{figure}
\begin{center}
\begin{tabular}{cc}
\includegraphics[scale=0.5]{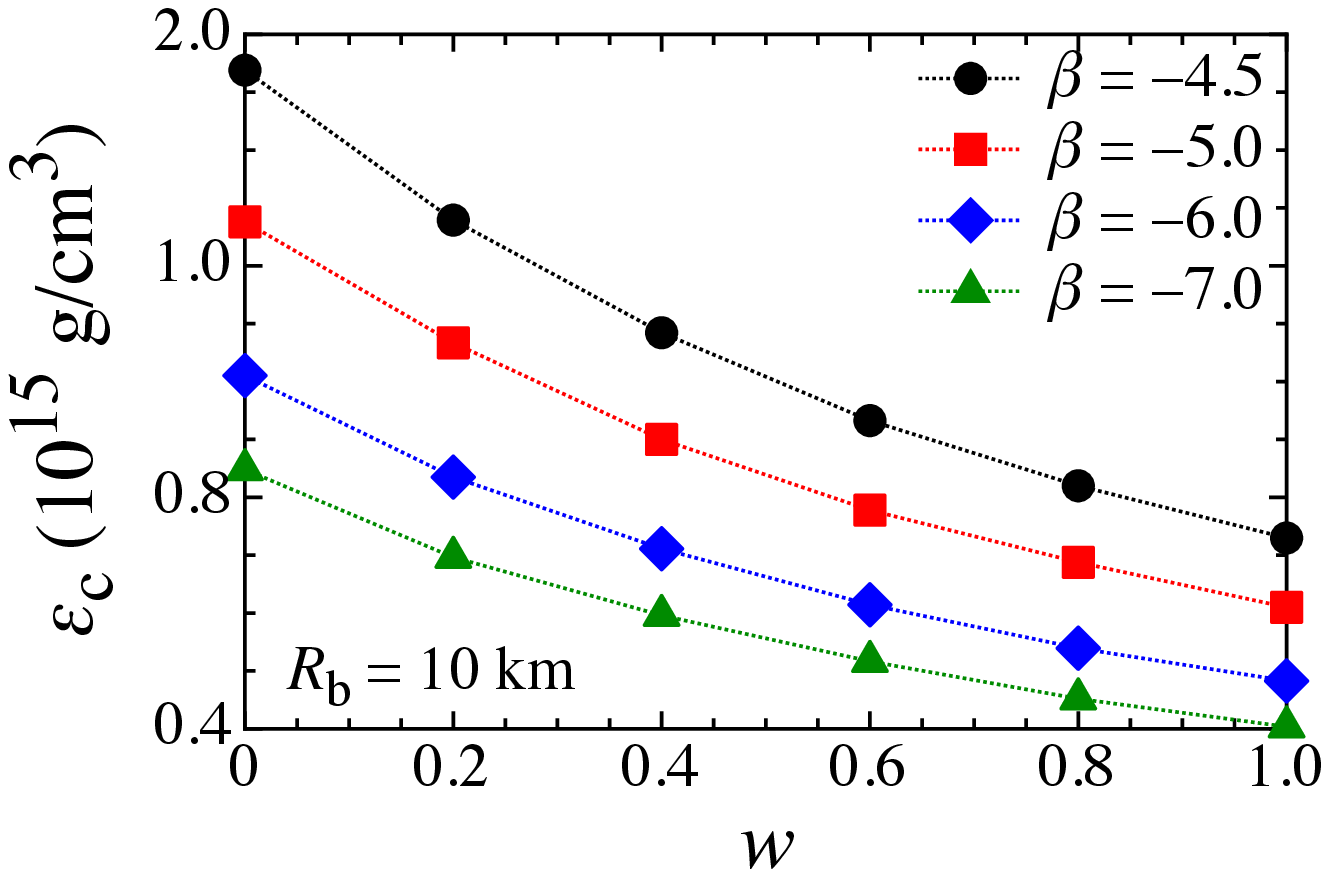} &
\includegraphics[scale=0.5]{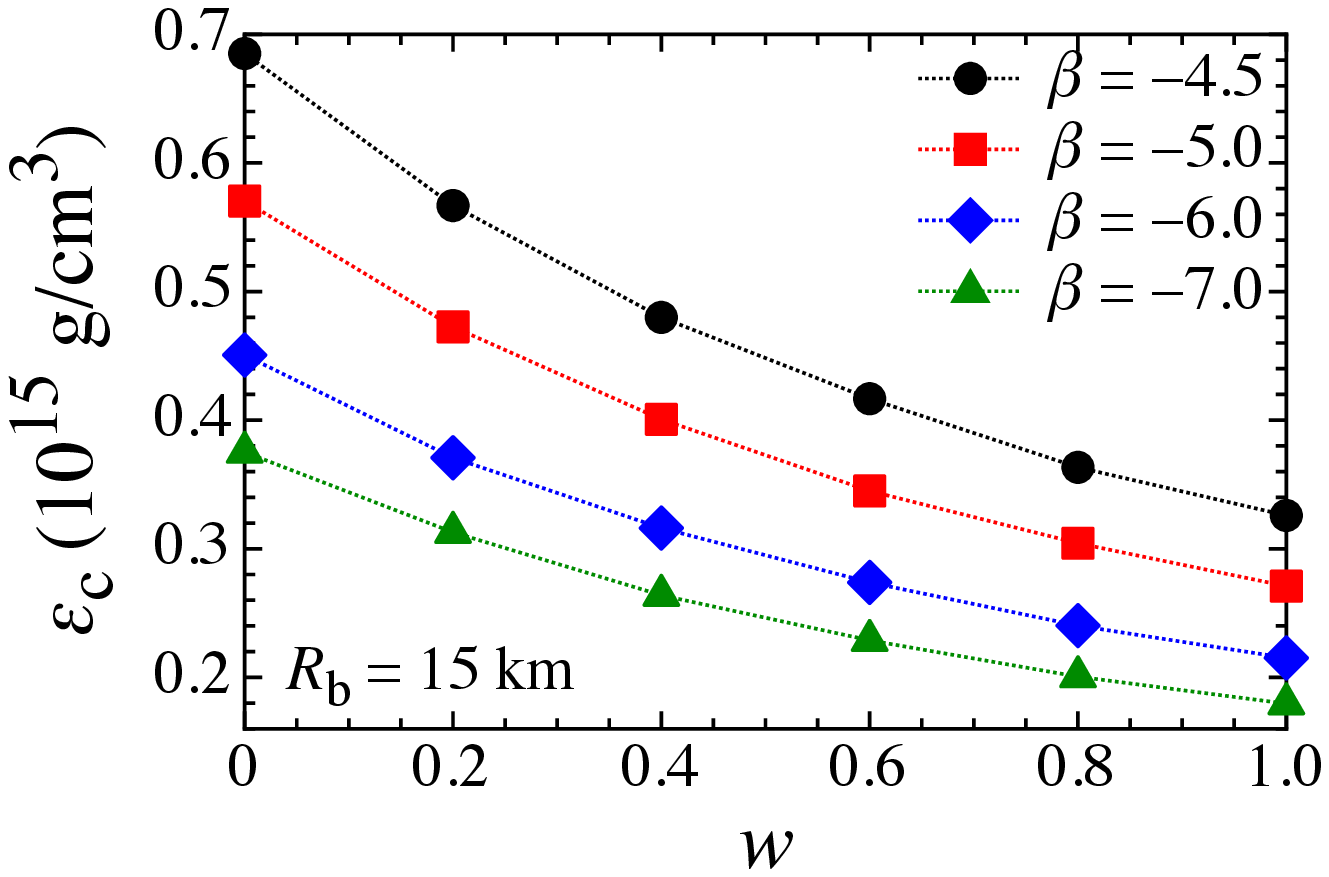} 
\end{tabular}
\end{center}
\caption{
Critical value of the central energy density, where the scalarization happens, is shown as a function of $w$ in Eq. (\ref{eq:er}) for various values of $\beta$, where the left and right panels correspond to the results with $R_b=10$ and 15 km.
}
\label{fig:weR10}
\end{figure}

Moreover, as mentioned above, we observe that the critical central energy density, for which the scalarization sets in, becomes smaller, for increasing $-\beta$ while it increases for decreasing $w$ (for fixed $-\beta$). These features can be viewed in  Fig. \ref{fig:weR10}, where the critical central energy density is plotted as function of the parameter $w$.

\begin{figure}
\begin{center}
\begin{tabular}{cc}
\includegraphics[scale=0.5]{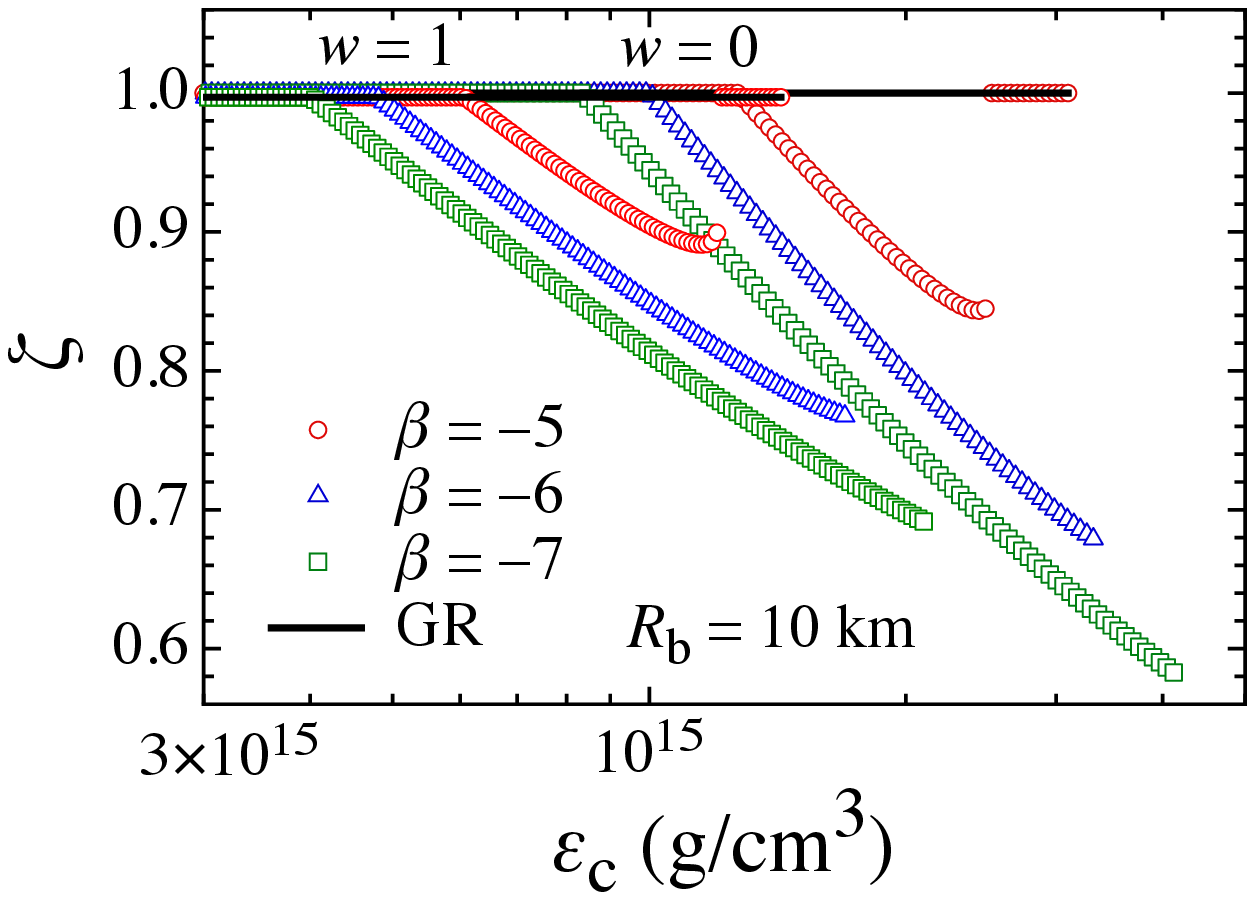} &
\includegraphics[scale=0.5]{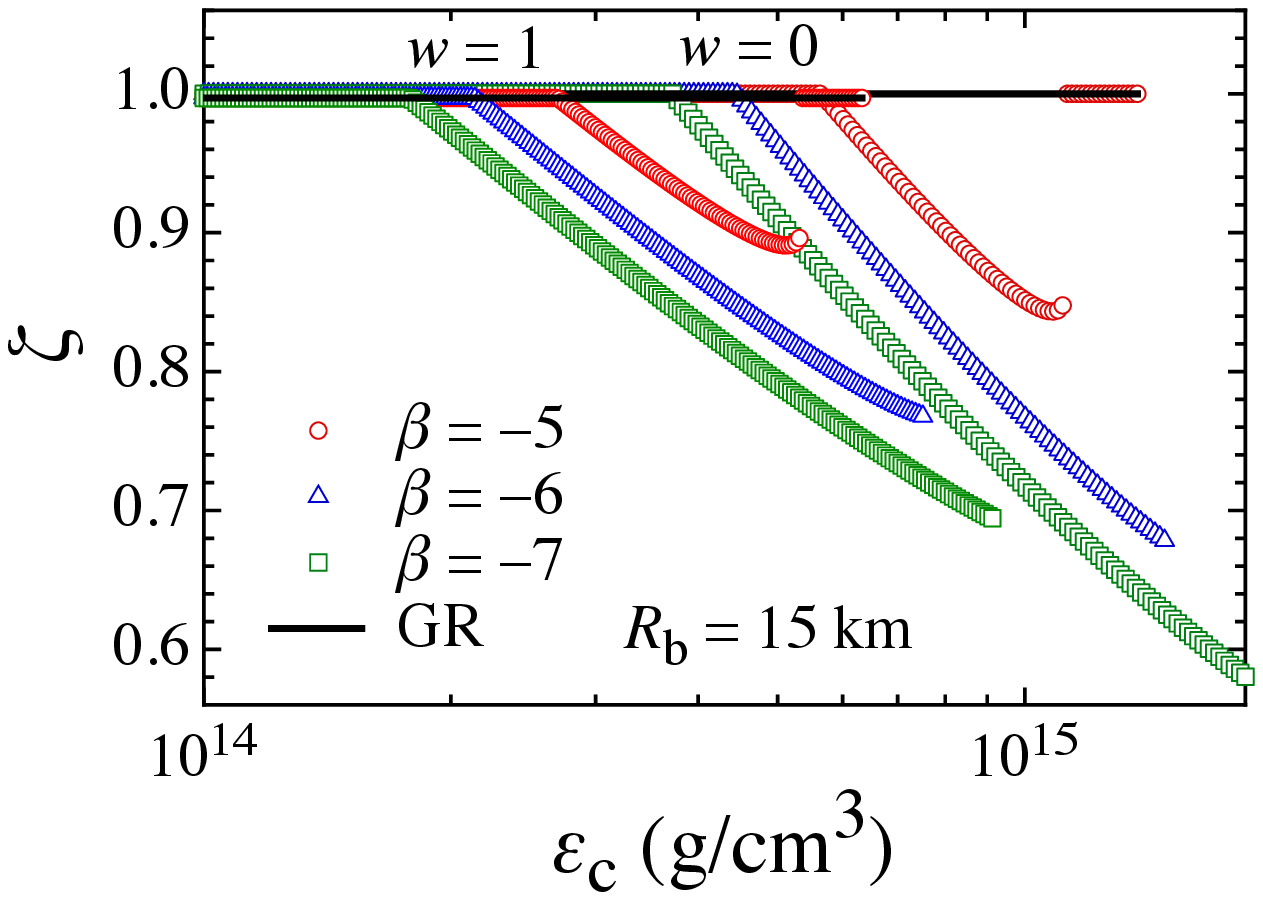} 
\end{tabular}
\end{center}
\caption{
The coefficient $\zeta$ in Eq. (\ref{eq:modi_mass}) is shown as a function of $\varepsilon_c$ for the three values of the coupling constants and for $w=0$ and 1. The left and right panels correspond to the cases with $R_b=10$ and 15 km.
}
\label{fig:ezeta}
\end{figure}

Finally, we {numerically} derive the mass formula modified in scalar-tensor gravity, such as
\begin{equation}
  M_{\rm ADM} = \zeta \times \frac{4\pi}{15}(2+3w) \varepsilon_c \label{eq:modi_mass} R^3 \, ,
\end{equation}
where $\zeta$ is a coefficient depending on the presence of the scalar field while for  $\zeta=1$ we recover general relativity \ie Eq. (\ref{eq:mass}). The  values of $\zeta$ for several cases are shown in Fig. \ref{fig:ezeta}. From this figure, one can observe that the mass of scalarized Tolman VII solution may be up to $\sim 35\%$ smaller (for $\beta=-7$) as compared to its value in general relativity.

\section{Conclusion}
\label{sec:VI}

In this paper, we systematically examine the compactness of neutron stars, based on the EOS defined by Eq. (\ref{eq:eos1}), and models based on the Tolman VII solutions in scalar-tensor gravity. First, we discussed the possible maximum compactness, by examining the compactness of  neutron stars with maximum mass in general relativity and for scalarized models  in scalar-tensor gravity with $\beta=-5$. The critical values were found by assuming that neutron stars become dynamically  unstable if $\partial M_{\rm ADM}/\partial R > 0$ in the $M_{\rm ADM} - R$ plane and the central density is larger than that for maximuma mass neutron stars of the sequence of models. This criterion was used both for scalarized and normal neutron stars. By adopting a simple EOS characterized by the nuclear saturation parameter $\eta$ in the lower density region and the sound velocity $v_s$ in higher density {(in this paper we adopt $v_s=1$ )}, we found that the maximum compactness in general relativity and in scalar-tensor gravity (assuming $\beta=-5$) is a linear function  of $\eta$. In this prescription of the neutron stars we derived that the maximum compactness for the general relativistic models is  $M_{\rm ADM}/R \sim 0.34$  while for the scalarized models  $M_{\rm ADM}/R \sim 0.29$ (assuming $\beta=-5$) for a range of plausible values of $\eta$. Next, we explored the possibility of constructing ultra-compact stars in scalar-tensor theory. For this search we adopted the analytic solution Tolman VII, which can cover a wide range of plausible EOSs with proper parametrization. {We numerically constructed the Tolman VII solutions in general relativity and in scalar-tensor theory by integrating the modified TOV equations.} We found that the compactness and mass of scalarized Tolman VII solutions are always smaller than that of the equivalent models in general relativity for the same set of parameters in the Tolman solution. Unlike the case in general relativity, scalarized ultra-compact stars can not be constructed in scalar-tensor gravity even for the  uniform density case.

\acknowledgments
This work was supported in part by Grant-in-Aid for Scientific Research (C) through Grant No. 17K05458.




\begin{thebibliography}{999}

\bibitem{Berti2015} 
   E. Berti et al., Class. Quant. Grav. {\bf 32}. 243001 (2015).

\bibitem{DP2003} 
   S. DeDeo and D. Psaltis, Phys. Rev. Lett. {\bf 90}, 141101 (2003).

\bibitem{SK2004} 
   H. Sotani and K. D. Kokkotas, Phys. Rev. D {\bf 70}, 084026 (2004); {\bf 71}, 124038 (2005).

\bibitem{Sotani2014} 
   H. Sotani, Phys. Rev. D {\bf 89}, 064031 (2014); {\bf 89}, 124037 (2014).

\bibitem[\protect\citeauthoryear{Damour \& Esposito-Far\`{e}se}{1993}]{DE1993}
   T.~Damour and G.~Esposito-Far\`{e}se, Phys. Rev. Lett. {\bf 70}, 2220 (1993).

\bibitem[\protect\citeauthoryear{Harada}{1998}]{Harada1998}
   T.~Harada, Phys. Rev. D {\bf 57}, 4802 (1998).

\bibitem[\protect\citeauthoryear{Doneva et al.}{2013}]{DYSK2013}
   D.~D.~Doneva, S.~S.~Yazadjiev, N.~Stergioulas, and K.~D.~Kokkotas, Phys. Rev. D {\bf 88}, 084060 (2013).

\bibitem[\protect\citeauthoryear{Barausse et al.}{2013}]{BPPL2013}
   E. Barausse, C. Palenzuela, M. Ponce, and L. Lehner, Phys. Rev. D {\bf 87}, 081506 (2013).

\bibitem[\protect\citeauthoryear{Palenzuela et al.}{2014}]{PBPL2014}
   C.~Palenzuela, E.~Barausse, M.~Ponce, and L.~Lehner, Phys. Rev. D {\bf 89}, 044024 (2014).

\bibitem[\protect\citeauthoryear{Taniguchi, Shibata, \& Buonanno}{2015}]{TSB2015}
   K.~Taniguchi, M.~Shibata, and A.~Buonanno, Phys. Rev. D {\bf 91}, 024033 (2015).

\bibitem[\protect\citeauthoryear{Freire et al.}{2012}]{Freire2012}
   P.~C.~C. Freire, N. Wex, G. Esposito-Far\`{e}se, J. P. W. Verbiest, M. Bailes, B. A. Jacoby, M. Kramer,
   I. H. Stairs, J. Antoniadis, and G. H. Janssen, Mon. Not. Roy. Astron. Soc. {\bf 423}, 3328 (2012).

\bibitem{shapiro-teukolsky}
   S. L. Shapiro and S. A. Teukolsky, in {\it Black Holes, White Dwarfs, and Neutron Stars} (Wiley-Interscience, 1983).

\bibitem[\protect\citeauthoryear{Demorest et al.}{2010}]{D2010}
   P.~B.~Demorest, T.~Pennucci, S.~M.~Ransom, M.~S.~E.~Roberts, and J.~W.~T.~Hessels, Nature {\bf 467}, 1081 (2010).

\bibitem[\protect\citeauthoryear{Antoniadis et al.}{2013}]{A2013}
   J. Antoniadis, et al., Science {\bf 340}, 1233232 (2013).

\bibitem{Abbott2017}
    B. P. Abbott, et al., Phys. Rev. Lett. {\bf 119}, 161101 (2017).

\bibitem{Bauswein2017} 
   A.~Bauswein, O.~Just, H.-T.~Janka, and N.~Stergioulas, Astrophys. J. Lett. {\bf 850}, L34 (2017).

\bibitem{Radice2017}
   D.~Radice, A.~Perego, F.~Zappa, and S.~Bernuzzi, Astrophys. J. Lett. {\bf 852}, L29 (2018).

\bibitem{Most2018}
   E.~R.~Most, L.~R.~Weih, L.~Rezzolla, and J.~ Schaffner-Bielich, arXiv:1803.00549

\bibitem{NICER}
   NICER project, https://www.nasa.gov/nicer

\bibitem{SM2017}
   H. Sotani and U. Miyamoto, Phys. Rev. D {\bf 96}, 104018 (2017).

\bibitem{S2017}
   H. Sotani, Phys. Rev. D {\bf 96}, 104010 (2017).

\bibitem{AK1998}
   N. Andersson and K. D. Kokkotas, Mon.\ Not.\ R. Astron.\ Soc.\ {\bf 299}, 1059 (1998).
 
 \bibitem{KS1992}
   K. D. Kokkotas and B. F. Schutz, Mon.\ Not.\ R. Astron.\ Soc.\  {\bf 255}, 119 (1992)

\bibitem{CF1991}
   S. Chandrasekhar and V. Ferrari, Proc. Royal Soc. Lond. A {\bf 434}, 449 (1991).

\bibitem{K1994} 
   K. D. Kokkotas, Mon.\ Not.\ R. Astron.\ Soc. {\bf 268}, 1015 (1994).

\bibitem{ultra-GR}
   Known examples of ultra-compact stars include the Tolman VII class of solutions in general relativity.

\bibitem{TKW98}
   T. Tsuchida, G. Kawamura, and K. Watanabe, Prog. Theor. Phys. {\bf 100}, 291 (1998).

\bibitem{TolmanVII} 
   R. C. Tolman, Phys. Rev. {\bf 55}, 364 (1939).

\bibitem{Moust2017} 
  Ch. C. Moustakidis, Gen. Rel. Gravit. {\bf 49}, 68 (2017).

\bibitem[\protect\citeauthoryear{Hartle}{1978}]{H1978}
   J.~B.~Hartle, Phys. Rep. {\bf 46}, 201 (1978).

\bibitem{LPMY1990}
   J. M. Lattimer, M. Prakash, D. Masak, and A. Yahil, Astrophys. J. {\bf 355}, 241 (1990).

\bibitem[\protect\citeauthoryear{Kalogera and Baym}{1996}]{KB1996}
   V. Kalogera and G. Baym, Astrophys. J. {\bf 470}, L61 (1996).

\bibitem[\protect\citeauthoryear{Koranda, Stergioulas, \&Friedman}{1997}]{KSF1997}
   S.~Koranda, N.~Stergioulas, and J.~L.~Friedman, Astrophys. J. {\bf 488}, 799 (1997).

\bibitem[\protect\citeauthoryear{Sotani}{2016}]{Sotani2016}   
   H. Sotani, Phys. Rev. C {\bf 95}, 025802 (2017).

\bibitem{SK2017} 
   H. Sotani and K. D. Kokkotas, Phys. Rev. D {\bf 95}, 044032 (2017).

\bibitem[\protect\citeauthoryear{Lattimer}{1981}]{L1981}
   J.~M.~Lattimer, Annu.\ Rev.\ Nucl.\ Part.\ Sci. {\bf 31}, 337 (1981).

\bibitem[\protect\citeauthoryear{Lattimer \& Lim}{2013}]{LL2013}
J.~M.~Lattimer and Y.~Lim, Astrophys. J.  {\bf 771}, 51 (2013).


\bibitem[\protect\citeauthoryear{Khan \& Margueron}{2013}]{Khan2013} 
   E.~Khan and J.~Margueron,  Phys. Rev. C {\bf 88}, 034319 (2013).

\bibitem[\protect\citeauthoryear{Newton et al.}{2014}]{Newton2014} 
   W.~G. Newton, J.~Hooker, M.~Gearheart, K. Murphy, D. H. Wen, F. J. Fattoyev, and B. A. Li, 
   Eur. Phys. J. A {\bf 50}, 41 (2014).

\bibitem[\protect\citeauthoryear{Sotani et al.}{2014}]{SIOO2014} 
   H.~Sotani, K.~Iida, K.~Oyamatsu, and A.~Ohnishi, Prog.\ Theor.\ Exp.\ Phys. {\bf 2014}, 051E01 (2014).

\bibitem[\protect\citeauthoryear{Silva, Sotani, \& Berti}{2016}]{SSB2016} 
   H.~O.~Silva, H.~Sotani, and E.~Berti, Mon. Not. R. Astron. Soc. {\bf 459}, 4378 (2016).

\bibitem[\protect\citeauthoryear{Oyamatsu \& Iida}{2003}]{OI2003}
   K.~Oyamatsu and K.~Iida, Prog.\ Theor.\ Phys. {\bf 109}, 631 (2003).

\bibitem[\protect\citeauthoryear{Oyamatsu \& Iida}{2007}]{OI2007}
   K.~Oyamatsu and K.~Iida, Phys.\ Rev.\ C {\bf 75}, 015801 (2007).

\bibitem[\protect\citeauthoryear{Lattimer \& Prakash}{2013}]{LP2001}
   J.~M.~Lattimer and M.~Prakash, Astrophys. J.  {\bf 550}, 426 (2001).

\bibitem{KRA2004}
   K.~D.~Kokkotas, J.~Ruoff, and N.~Andersson, Phys. Rev. D {\bf 70}, 043003 (2004).

\bibitem{Card2008}
   V.~Cardoso, P.~Pani, M.~Cadoni, M.~Cavaglia Phys. Rev. D {\bf 77}, 124044 (2008).

\bibitem{CP2017}
   V.~Cardoso and P.~Pani, Nat. Astron. {\bf 1}, 586 (2017).

\bibitem{Abedi2017}
   J.~Abedi, H.~Dykaar, and N.~Afshordi, Phys. Rev. D {\bf 96}, 082004 (2017).

\bibitem{MVK2017}
   A.~Maselli, S.~H.~V\"olkel, and K.~D.~Kokkotas, Phys. Rev. D {\bf 96}, 064045 (2017).







\end{thebibliography}
\end{document}